\documentclass[conference]{IEEEtran}
\IEEEoverridecommandlockouts
\usepackage{cite}
\usepackage{amsmath,amssymb,amsfonts}
\usepackage{stfloats}
\usepackage{algorithmic}
\usepackage{graphicx}
\usepackage{textcomp}
\usepackage{xcolor}
\usepackage{multirow}
\def\BibTeX{{\rm B\kern-.05em{\sc i\kern-.025em b}\kern-.08em
    T\kern-.1667em\lower.7ex\hbox{E}\kern-.125emX}}
\begin{document}

\title{An End-to-End Deep Learning Approach for \\
Epileptic Seizure Prediction
}

\author{\IEEEauthorblockN{Yankun Xu$^{1}$, Jie Yang$^{1}$, Shiqi Zhao$^{1}$, Hemmings Wu$^{2}$, and Mohamad Sawan$^{1}$, \IEEEmembership{Fellow, IEEE}}
\IEEEauthorblockA{$^{1}$CenBRAIN, School of Engineering, Westlake University, Hangzhou, Zhejiang, China 310024 \\
$^{2}$Department of Neurosurgery, Zhejiang University School of Medicine Second Affiliated Hospital, \\
Hangzhou, Zhejiang, China 310009}
xuyankun@westlake.edu.cn\\

}

\maketitle
\begin{abstract}
An accurate seizure prediction system enables early warnings before seizure onset of epileptic patients. It is extremely important for drug-refractory patients. Conventional seizure prediction works usually rely on features extracted from Electroencephalography (EEG) recordings and classification algorithms such as regression or support vector machine (SVM) to locate the short time before seizure onset. However, such methods cannot achieve high-accuracy prediction due to information loss of the hand-crafted features and the limited classification ability of regression and SVM algorithms. We propose an end-to-end deep learning solution using a convolutional neural network (CNN) in this paper. One and two dimensional kernels are adopted in the early- and late-stage convolution and max-pooling layers, respectively. The proposed CNN model is evaluated on Kaggle intracranial and CHB-MIT scalp EEG datasets. Overall sensitivity, false prediction rate, and area under receiver operating characteristic curve reaches 93.5\%, 0.063/h, 0.981 and 98.8\%, 0.074/h, 0.988 on two datasets respectively. Comparison with state-of-the-art works indicates that the proposed model achieves exceeding prediction performance.
\end{abstract}

\quad


\section{Introduction}
Epilepsy is one of the most common neurological diseases worldwide, most patients with epilepsy are treated with long-term drug therapy, but approximately a third of patients are drug refractory\cite{kwan2010definition,granata2009management,assi2017refractory}. Patient with drug-resistant epilepsy can benefit from intervenes in advance of the seizure onset to reduce the possibility of severe comorbidities and injuries\cite{canevini2010relationship,tellez2007psychiatric}. Therefore, a closed-loop system with accurate seizure forecasting capacity plays an important role in improving the life quality of patients and lower cost of healthcare resources\cite{nagaraj2015future,assi2017towards}. However, seizure prediction faces many difficulties and challenges, many previous studies even stated that epileptic seizure is hardly to predict \cite{golestani2014can,kalitzin2014predicting,di2007switching}.

Electroencephalography (EEG) is a kind of electrophysiological technique used to record the electrical activity of the brain. Scalp and intracranial EEG are usually utilized for epileptic seizure monitor. EEG recordings from epileptic patients can be defined as several intervals: ictal (seizure onset), preictal (short time before the seizure), postictal (short time after the seizure), and interictal (between seizures but excluding preictal and postictal). The prediction task is to identify preictal state that can be differentiated from the interictal, ictal, and postictal states, and the major challenge in seizure prediction is to differentiate between the preictal and interictal states\cite{hussein2019human,zhang2015seizure,aarabi2009eeg}.

Over the past decade, several research activities were based on EEG recordings to predict seizure\cite{eberlein2018convolutional,khan2017focal,truong2018convolutional}. Most researchers make use of manually selected features within the frequency domain to make prediction\cite{assi2017functional,assi2015hybrid,gagliano2018bilateral}. However, there are some limitations for hand crafted feature engineering. Firstly it cannot be generalized for EEG recording acquired from various devices. Secondly some important information might be lost during the features extraction process. Thirdly feature extraction operation increases computational complexity for real-time applications. Hence recent work relies on deep learning to automatically learn features from EEG recordings\cite{eberlein2018convolutional,avcu2019seizure,tsiouris2018long}. However, these works take EEG as conventional 2-dimensional (2D) signals such as image and fail to consider its unique spatial-temporal characteristics\cite{freeman2012imaging,mognon2011adjust}.

In this study, an end-to-end patient-specific approach using convolutional neural network (CNN) is proposed to address seizure prediction challenges\cite{lecun1995convolutional}.
The proposed network adopts 1-dimensional (1D) kernel in the early-stage convolution and max-pooling layer to make use of the EEG signal redundancy in the time domain and preserve information in the spatial domain\cite{ranzato2008sparse,jarrett2009best}. 2D kernels are used in the late-stage to combine the information from multiple channels to make high accurate prediction.

The remaining parts of this paper are organized as follows. Section \ref{2} describes the two targeted datasets used for developing the achieved seizure prediction algorithm. Section \ref{3} introduces the proposed model architecture and its implementation. Performance evaluation and comparison with the state-of-the-art are carried out in Section \ref{4}. The last section discusses and concludes our contribution in this paper.

\section{Benchmark EEG Datasets} \label{2}

In this work, our model is implemented and evaluated on two widely used benchmark EEG datasets.
\begin{itemize}
\item Intracranial EEG dataset: The Kaggle dataset consists of 5 dogs and 2 human patients\cite{brinkmann2016crowdsourcing}. All five dogs are sampled at 400Hz, four of them are recorded with 16 electrodes and one with 15 electrodes. As to human patients, their sample rate is 5000Hz, one is recorded with 15 electrodes and one with 24 electrodes.
\item Scalp EEG dataset: The CHB-MIT dataset collected 23 subjects from 22 patients for seizure detection purpose originally\cite{goldberger2000physiobank,shoeb2009application}. There are 637 recordings in total in the dataset, each recording may contain none or several seizures. All subjects are recorded at the same 256Hz sample rate, but in term of electrodes, there are different signal acquisition settings among patients. 15 subjects are recorded from fixed 23-electrode configuration, while one or several changes of electrodes' setting are implemented for the remaining subjects. 
\end{itemize}

In this study, only lead seizures, which are defined as seizures occurring at least 4h after previous seizures, are considered\cite{hussein2019human}. Furthermore, in term of Kaggle dataset, two human patients are excluded since they are recorded with significantly different signal acquisition approach.
As to CHB-MIT dataset, we only consider subjects with fixed electrodes configuration, and no less than 3 lead seizures, so that there are only 7 subjects suitable for experiments. 

\begin{figure}[t]
\centering
\includegraphics[width=8.5cm]{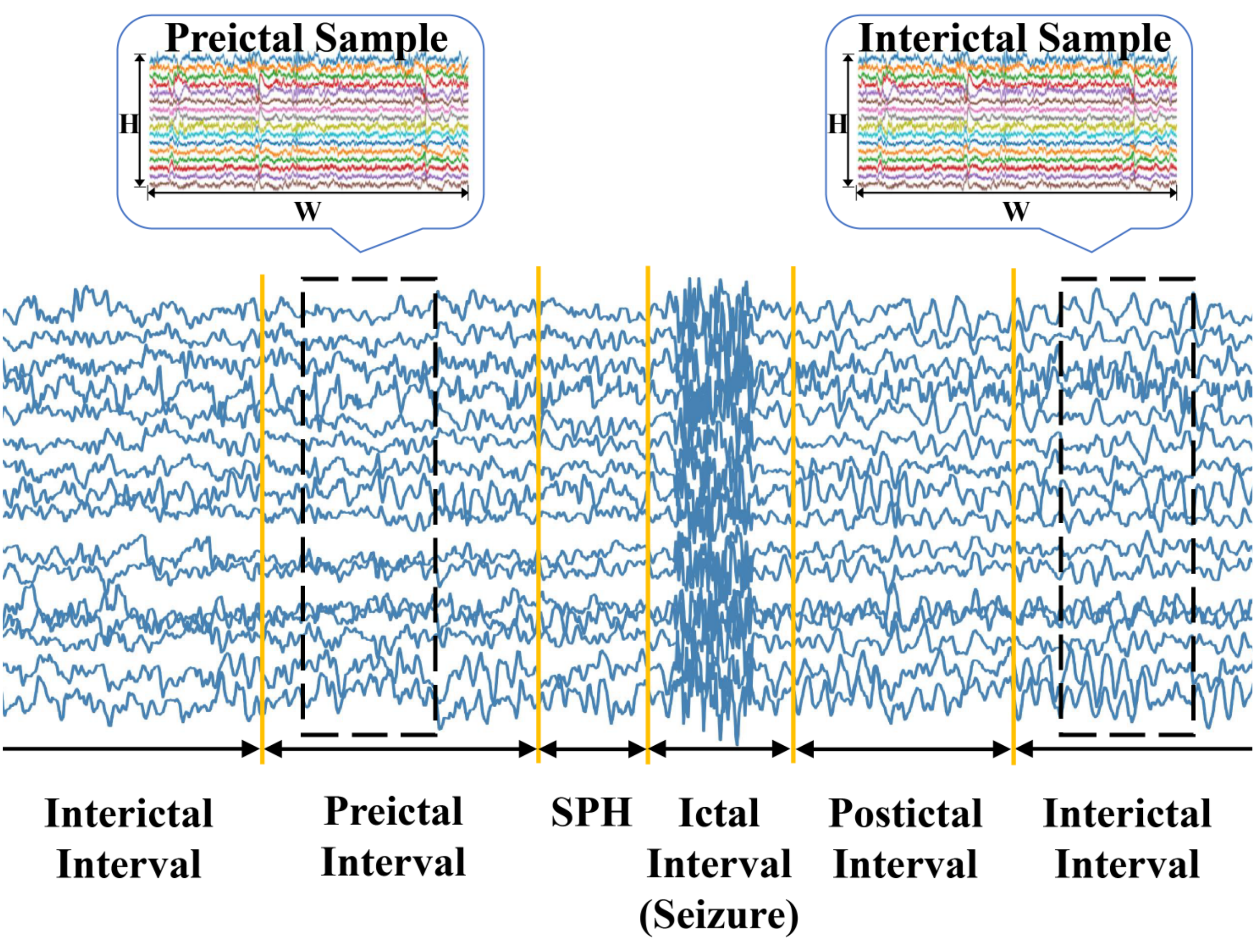}
\caption{An example of EEG recording from epileptic patient. Different timing definitions are shown. Seizure prediction horizon (SPH) is a interval between preictal interval and seizure onset. Except for intervals of preictal, postictal, ictal, and SPH, the remaining recording belongs to interictal interval.}
\label{period}
\end{figure}

\begin{figure*}[t]
\centering
\includegraphics[width=\textwidth]{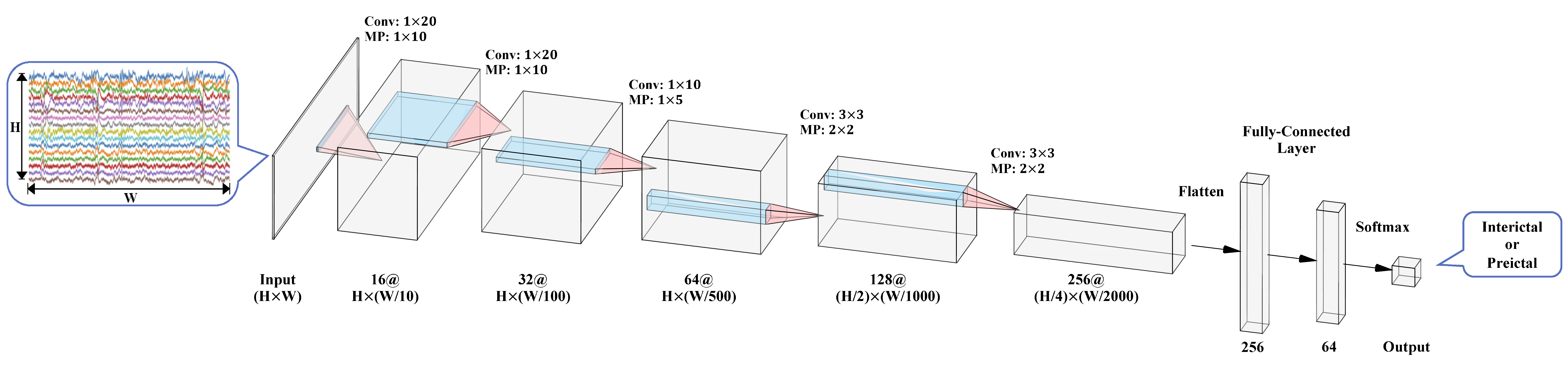}
\caption{Architecture of proposed model. Conv and MP stands for convolution and max-pooling operation respectively. When we feed the input with shape Height$\times$Width (H$\times$W) to this model, output with softmax function will generate the probability of input belonging to preictal state. This model consists of 5 convolution layers with ReLU activation function and 2 fully-connected (FC) layers with sigmoid activation function. There is no batch normalization and regularization in this model. The description on the top of connection between convolution blocks shows the size of convolution and max-pooling kernel, and the description at the bottom of each convolution block shows the number of kernels and the shape of each layer. The two FC layers contains 256 and 64 nodes respectively.}
\label{model}
\end{figure*}


Figure \ref{period} shows timing definitions of different EEG signal contents. In seizure prediction task, a short interval between end of preictal interval and seizure onset is defined as seizure prediction horizon (SPH). However, preictal interval length (PIL) and SPH are two empirical parameters determining which interval of recording is chosen for experiments. In term of Kaggle dataset, 1h PIL and 5min SPH were already set during dataset preparation, but for CHB-MIT, 30min PIL and 5min SPH are chosen for fair comparison to other state-of-the-art works.

In this study, two categorized samples are extracted from interictal and preictal intervals respectively with fixed 20s time window, as shown by dashed box in Fig. \ref{period}. Then the width of input sample is $20s \times sample\_rate$, and height is the number of recording channels. Hence the shape of input from two datasets is 16 (or 15) $\times$ 8000, and $23\times5120$ respectively. 16 (or 15) and 23 are the numbers of their channels, and 8000 and 5120 are the numbers of selected data points which equal to the multiplication results of the time window (20s) and their sample rates (400Hz, 256Hz). However, raw
EEG recording of epileptic patient contains much longer interictal interval than preictal interval, imbalanced
training sample problem may cause trained model to achieve poor performance\cite{barandela2004imbalanced,japkowicz2002class}. To overcome it, we extract interictal samples from EEG recordings without any overlapping, but preictal samples are extracted with 5s overlapping. Thus, according to raw EEG recordings of various cases, different numbers of interictal and preictal samples are extracted.

\section{Proposed CNN Model} \label{3}

Unlike image data that have the vertical and horizontal resolutions in the same magnitude, the formulated EEG signal samples have thousands of elements in the time axis but only few in the channel axis. This unique feature distinguishes EEG from image. The large redundancy in both vertical and horizontal directions of a 2D image makes 2D convolution perfectly suitable for extracting hidden features from image\cite{pan2000image}. However, EEG signal is not redundant in the channel axis, because each channel records different area of the brain. Based on the above consideration, the proposed model adopts 1D and 2D kernels for convolution and max-pooling operation in early- and late-stage layers, respectively. The initial 1D convolution kernels are able to learn features within time domain and 1D max-pooling kernels help keep only significant information. The remaining convolution and max-pooling layers utilize 2D kernels to learn features within both spatial and time domains. 
If only 2D kernels are adopted in the network, the channel dimension will soon shrink which results in a shallow network structure.


Our CNN model is shown in Fig. \ref{model}. Except for input and output layers, proposed model consists of 5 convolution layers and 2 fully-connected (FC) layers. Each convolution layer is followed by a rectified linear unit (ReLU) activation function and one max-pooling layer\cite{nair2010rectified}.
The size of the convolution kernels and max-pooling kernels are also shown in Fig. \ref{model}. Convolution and max-pooling kernels with size of $1\times20$ and $1\times10$ are adopted for first two convolution blocks, for the following third convolution block, convolution kernel with size of $1\times10$ and max-pooling kernel with size of $1\times5$ are used.

For the remaining two convolution layers, convolution kernel with size of $3\times3$ and max-pooling kernel with size of $2\times2$ are implemented.
The number of kernels for the five convolution layers is 16, 32, 64, 128, 256 respectively.
The two FC layers with 256 and 64 nodes respectively are followed by a sigmoid activation function. The output layer uses Softmax activation function for binary classification and binary cross entropy is used as loss function\cite{rubinstein2013cross}.
There is only one dropout layer with 0.5 dropout rate settled between the last convolution layer and the first FC layer\cite{srivastava2014dropout}.

We adopted Adam optimizer for loss minimization with learning rate, $\beta_{1}$ and $\beta_{2}$ of 1e-5, 0.9, 0.999 respectively\cite{kingma2014adam}. Even though we extracted preictal samples with overlapping, the number of training sample is still imbalanced. To overcome this issue, we randomly feed equal number of interictal and preictal samples from training set to model for each training epoch. Training of each case is stopped after 100 epochs. Our model is implemented in Python 3.6 using Keras 2.2 with Tensorflow 1.13 backend on single NVIDIA 2080Ti GPU\cite{abadi2016tensorflow}. For one epoch, 6400 samples are trained, and the training process takes 270s and 30s on Kaggle and CHB-MIT dataset respectively.

\begin{table}[t]
\renewcommand\arraystretch{1.2}
\caption{ }
\begin{center}

\begin{tabular}{c c c c c}
\hline\hline
\multirow{3}{*}{\textbf{\scriptsize{Subject}}} & \textbf{\scriptsize{No. of}} & \multirow{3}{*}{\textbf{\scriptsize{Sensitivity(\%)}}} & \multirow{3}{*}{\textbf{\scriptsize{FPR(/h)}}} & \multirow{3}{*}{\textbf{\scriptsize{AUC}}} \\
& \textbf{\scriptsize{LS/}} &  &  & \\
& \textbf{\scriptsize{seizures}} &  &  & \\
\hline
Dog 1 & 4/4 & $90.6 \pm 5.2$ & $0.053 \pm 0.022$ & $0.983 \pm 0.002$\\
Dog 2 & 7/7 & $96.8 \pm 1.5$ & $0.028 \pm 0.012$ & $0.996 \pm 0.001$\\
Dog 3 & 12/12 & $93.1 \pm 4.1$ & $0.062 \pm 0.030$ & $0.990 \pm 0.000$\\
Dog 4 & 16/16 & $88.7 \pm 4.6$ & $0.148 \pm 0.028$ & $0.941 \pm 0.005$\\
Dog 5 & 5/5 & $98.6 \pm 0.9$ & $0.022 \pm 0.007$ & $0.996 \pm 0.001$\\
\hline
Overall & 44/44 & $93.5 \pm 3.3$ & $0.063 \pm 0.020$ & $0.981 \pm 0.002$\\
\hline\hline
\end{tabular}

        
\label{tab1}
\end{center}
\end{table}

\begin{table}[t]
\renewcommand\arraystretch{1.2}
\caption{ }
\begin{center}
\begin{tabular}{c c c c c}
\hline\hline
\multirow{3}{*}{\textbf{\scriptsize{Subject}}} & \textbf{\scriptsize{No. of}} & \multirow{3}{*}{\textbf{\scriptsize{Sensitivity(\%)}}} & \multirow{3}{*}{\textbf{\scriptsize{FPR(/h)}}} & \multirow{3}{*}{\textbf{\scriptsize{AUC}}} \\
& \textbf{\scriptsize{LS/}} &  &  & \\
& \textbf{\scriptsize{seizures}} &  &  & \\
\hline
chb01 & 3/7 & $100.0 \pm 0.0$ & $0.001 \pm 0.001$ & $1.000 \pm 0.000$\\
chb05 & 3/5 & $99.7 \pm 0.4$ & $0.072 \pm 0.019$ & $0.993 \pm 0.001$\\
chb06 & 6/10 & $96.0 \pm 1.1$ & $0.138 \pm 0.023$ & $0.967 \pm 0.002$\\
chb08 & 3/5 & $99.9 \pm 0.2$ & $0.027 \pm 0.012$ & $0.998 \pm 0.001$\\
chb10 & 5/7 & $97.9 \pm 1.1$ & $0.095 \pm 0.043$ & $0.985 \pm 0.006$\\
chb14 & 4/8 & $98.9 \pm 1.0$ & $0.109 \pm 0.017$ & $0.983 \pm 0.005$\\
chb22 & 3/3 & $99.5 \pm 0.4$ & $0.078 \pm 0.023$ & $0.992 \pm 0.002$\\
\hline
Overall & 27/45 & $98.8 \pm 0.6$ & $0.074 \pm 0.020$ & $0.988 \pm 0.003$\\
\hline\hline
\end{tabular}

\qquad
\newline
\leftline{\text{\scriptsize{LS: Lead Seizures; FPR: False Prediction Rate; AUC: Area Under Curve.}}}

\label{tab2}
\end{center}
\end{table}

\section{Results} \label{4}
We first evaluate our model with standard metrics and then compare it with two other works that achieves state-of-the-art performance. Sensitivity, false prediction rate (FPR), receiver operating characteristic (ROC) curve, and area under curve (AUC) are evaluated in this study\cite{zweig1993receiver,hanley1982meaning}. For each subject, we spilt 20\% interictal and preictal samples for validation purpose, and the remaining is used for training the model. 10 independent runs with different initializers are implemented to generate mean and standard deviation of each metric in order to evaluate the robustness of model, and the validation set is chosen randomly for each run.

Tables \ref{tab1} and \ref{tab2} show results including subject information and measured metrics for the two datasets respectively. We take an average of each metric as an overall measurement.

\begin{figure}[t]
\centering
\includegraphics[width=9cm]{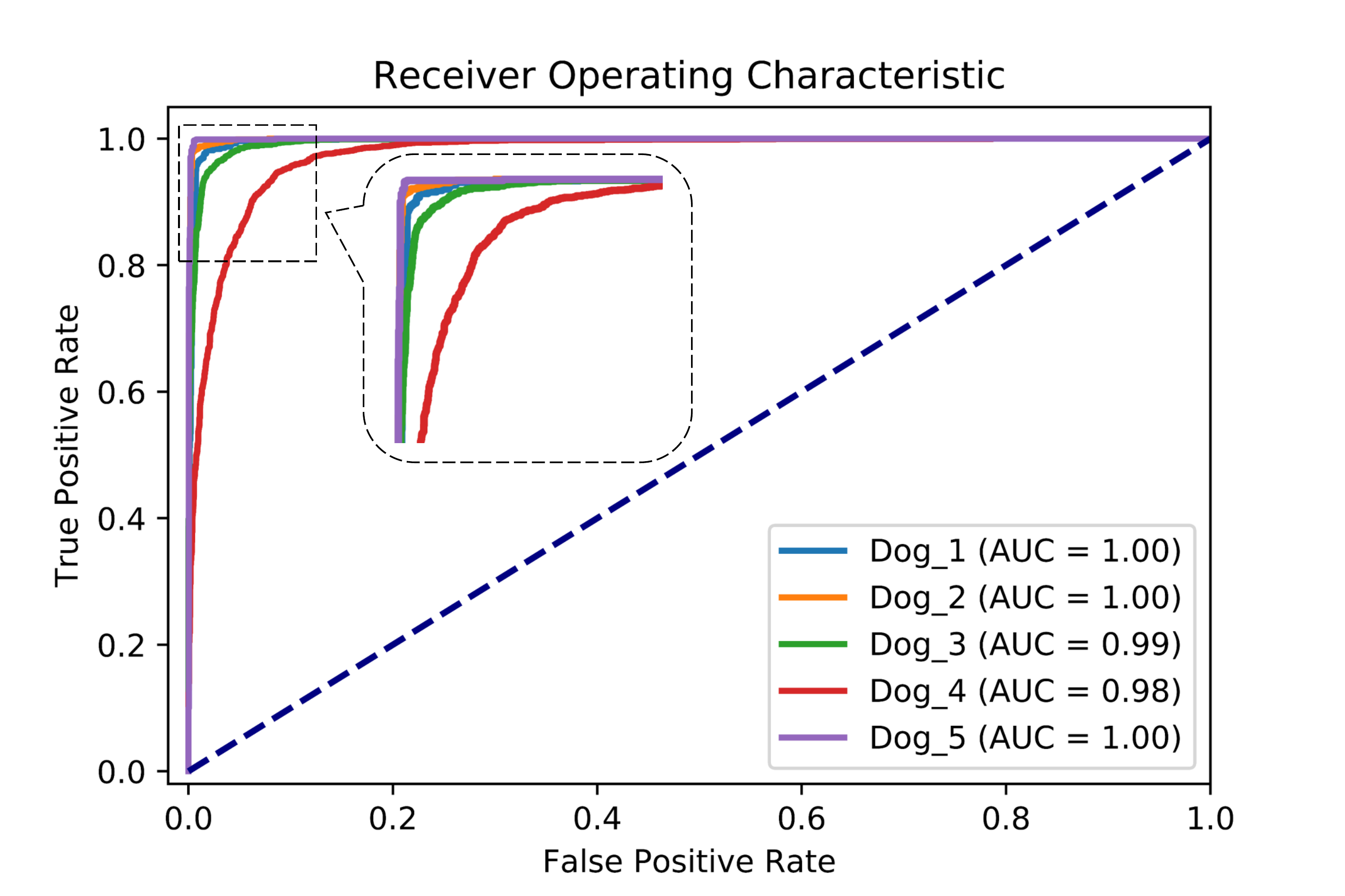}
\caption{ROC curves including AUC of all dog subjects from Kaggle dataset. The dash line represents the performance of random predictor. For each subject, only the best performance from 10 runs is shown.}
\label{ROC-1}
\end{figure}

\begin{figure}[t]
\centering
\includegraphics[width=9cm]{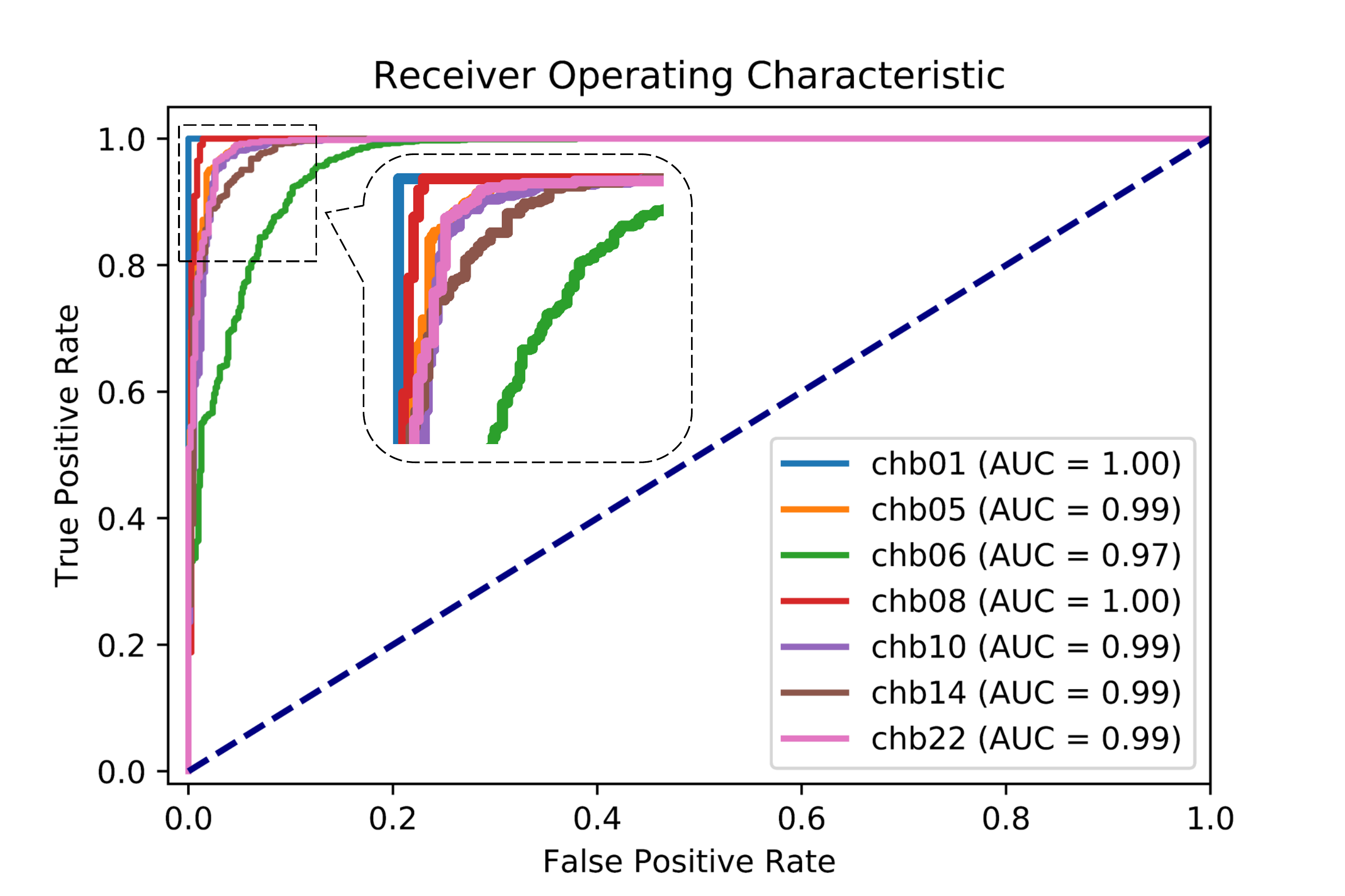}
\caption{ROC curves including AUC of all patients from CHB-MIT dataset. The dash line represents the performance of random predictor. For each subject, only the best performance from 10 runs is shown.}
\label{ROC-2}
\end{figure}

Performance of proposed model on Kaggle dataset reaches average 93.5\% sensitivity, 0.063 FPR, and 0.981 AUC score. The overall standard deviation of three metrics is 3.3\%, 0.02/h and 0.002, which indicates the model performance is quite robust on this dataset.
For CHB-MIT dataset, overall sensitivity, FPR and AUC reach 98.8\%, 0.074/h, and 0.988 respectively. Among all patients, four of them achieves very good results (reach $\geq99\%$ sensitivity and $\geq 0.99$ AUC at same time), but the subject \textit{chb01} reaches highest performance with 100\% sensitivity, 0 FPR and 1 AUC. Low standard deviation is also achieved on this dataset. From Figs. \ref{ROC-1} and \ref{ROC-2}, ROC curves with their AUC from each case shows that our model has good capacity to separate preictal and interictal samples of the two datasets.

We compare our model with two other state-of-the-art works. Some comparison results are summarized in Tables \ref{com1} and \ref{com2}. Table \ref{com1} shows comparisons of performance on Kaggle dataset. Truong et al. made use of feature extracted with short-time Fourier transform, then trained a CNN model to classify, except for the subject \textit{Dog 2}, they achieved obviously worse performance on the remaining subjects\cite{truong2018convolutional}. Eberlein et al. also utilized CNN to process raw EEG signal, two of four dogs only reach around 0.8 AUC, and overall AUC is less than 0.9, which is lower than AUC achieved from our model, while our model is able to achieve much higher AUC\cite{eberlein2018convolutional}.

Table \ref{com2} shows comparison with Truong's and Khan's works on the CHB-MIT dataset. Khan et al. combined wavelet transform and CNN to identify preictal interval\cite{khan2017focal}. Over the four patients listed by the two works, sensitivity is less than 90\%. All five patients from their work reach over 0.85 AUC, where \textit{chb05} achieves the highest AUC of 0.988, which is still worse than this work.

\begin{table}[t]
\renewcommand\arraystretch{1.2}
\caption{ }
\begin{center}

\begin{tabular}{c | c | c | c | c | c | c}
\hline
  & \multicolumn{2}{|c|}{Truong et al. [1]} & \multicolumn{2}{|c|}{Eberlein et al. [2]} & \multicolumn{2}{|c}{\textbf{This work}} \\
 & \multicolumn{2}{|c|}{(STFT + CNN)} & \multicolumn{2}{|c|}{(Raw + CNN)} & \multicolumn{2}{|c}{(Raw + CNN)}\\
\cline{2-7}
& SEN(\%) & AUC & SEN(\%) & AUC & SEN(\%) & AUC\\
\hline

Dog 1  & 50 & -- & -- & 0.798 & \textbf{90.6}& \textbf{0.983}\\
Dog 2  & 100 & -- & --  & 0.812 & \textbf{96.8}& \textbf{0.996}\\
Dog 3  & 58.3 & -- & -- & 0.844 & \textbf{93.1}& \textbf{0.990}\\
Dog 4  & 78.6 & -- & --  & 0.919 & \textbf{88.7}& \textbf{0.941}\\
Dog 5 & 80 & -- & -- & -- & \textbf{98.6}& \textbf{0.996}\\
\hline
\end{tabular}

\qquad


\label{com1}
\end{center}
\end{table}

\begin{table}[t]
\renewcommand\arraystretch{1.2}
\caption{ }
\begin{center}

\begin{tabular}{c | c | c | c | c | c | c}
\hline
  & \multicolumn{2}{|c|}{Truong et al. [1]} & \multicolumn{2}{|c|}{Khan et al. [3]} & \multicolumn{2}{|c}{\textbf{This work}} \\
 & \multicolumn{2}{|c|}{(STFT + CNN)} & \multicolumn{2}{|c|}{(Wavelet + CNN)} & \multicolumn{2}{|c}{(Raw + CNN)}\\
\cline{2-7}
& SEN(\%) & AUC & SEN(\%) & AUC & SEN(\%) & AUC\\
\hline

chb01 & 85.7 & -- & -- & 0.943& \textbf{100} & \textbf{1.000}\\
chb05 & 80.0& -- & -- & 0.988 & \textbf{99.7} & \textbf{0.993}\\
chb08 & --& --& -- & 0.921 & \textbf{99.9} & \textbf{0.998}\\
chb10 & 33.3& --& -- & 0.855 & \textbf{97.9} & \textbf{0.985}\\
chb14 & 80.0& --& -- & -- & \textbf{98.9} & \textbf{0.983}\\
chb22 & --& --& -- & 0.877 & \textbf{99.5} & \textbf{0.992}\\

\hline
\end{tabular}

\qquad

\leftline{\text{\scriptsize{STFT: Short-Time Fourier Transform; CNN: Convolution Neural Network}}}

\leftline{\text{\scriptsize{SEN: Sensitivity; AUC: Area Under Curve.}}}

\label{com2}
\end{center}
\end{table}

\section{Conclusion} \label{5}

We described an end-to-end CNN architecture for seizure prediction in this paper. Instead of using more commonly seen features from the frequency domain, raw EEG signals are used as input of the CNN model. Taking into account the unique characteristic of the EEG signal, the proposed architecture utilizes 1D convolution and max-pooling kernel to make use of the EEG signal redundancy in the time axis and preserve information in the channel axis for early-stage operation. Experimental results show that the proposed architecture achieves high sensitivity, AUC score and lower FPR on widely adopted benchmark datasets. Comparison result indicates that the model outperforms state-of-the-art works.

In addition, the use of raw signals allowed to reduce the complexity of data processing, which is expected to reduce the execution time, reduce the power consumption and shrink the silicon area in the projected hardware oriented implementation.

\section*{Acknowledgements}
The authors would like to acknowledge start-up funds from Westlake University to the Cutting-Edge Net of Biomedical Research and INnovation (CenBRAIN) to support this project. 

\clearpage
{
\bibliographystyle{IEEEtran}
\bibliography{egbib}
}

\end{document}